# DeepSeg: Deep Neural Network Framework for Automatic Brain Tumor Segmentation using Magnetic Resonance FLAIR Images

Ramy A. Zeineldin[1]*  · Mohamed E Karar[2]  · Jan Coburger[3]  · Christian R. Wirtz[3]  · Oliver Burgert[1]

[1]*Research Group Computer Assisted Medicine (CaMed), Reutlingen University, 72762 Reutlingen, Germany*
[2]*Faculty of Electronic Engineering (FEE), Menoufia University, 32952 Menouf, Egypt*
[3]*Department of Neurosurgery, University of Ulm, 89312 Günzburg, Germany*

**Abstract**

**Purpose** Gliomas are the most common and aggressive type of brain tumors due to their infiltrative nature and rapid progression. The process of distinguishing tumor boundaries from healthy cells is still a challenging task in the clinical routine. Fluid-Attenuated Inversion Recovery (FLAIR) MRI modality can provide the physician with information about tumor infiltration. Therefore, this paper proposes a new generic deep learning architecture; namely DeepSeg for fully automated detection and segmentation of the brain lesion using FLAIR MRI data.

**Methods** The developed DeepSeg is a modular decoupling framework. It consists of two connected core parts based on an encoding and decoding relationship. The encoder part is a convolutional neural network (CNN) responsible for spatial information extraction. The resulting semantic map is inserted into the decoder part to get the full resolution probability map. Based on modified U-Net architecture, different CNN models such as Residual Neural Network (ResNet), Dense Convolutional Network (DenseNet), and NASNet have been utilized in this study.

**Results** The proposed deep learning architectures have been successfully tested and evaluated on-line based on MRI datasets of Brain Tumor Segmentation (BraTS 2019) challenge, including s336 cases as training data and 125 cases for validation data. The dice and Hausdorff distance scores of obtained segmentation results are about 0.81 to 0.84 and 9.8 to 19.7 correspondingly.

**Conclusion** This study showed successful feasibility and comparative performance of applying different deep learning models in a new DeepSeg framework for automated brain tumor segmentation in FLAIR MR images. The proposed DeepSeg is open-source and freely available at https://github.com/razeineldin/DeepSeg/.

**Keywords** Brain Tumor . Computer-aided diagnosis . Convolutional neural networks. Deep learning.

## Introduction

Brain tumors are one of the leading causes of death for cancer patients, especially children, and young people. The American Cancer Society reported that 23,820 new brain cancer cases in the United States were discovered in 2019 [1]. Brain tumors can be categorized into two types as follow: Primary brain tumors that originate in the brain cells, and secondary brain tumors developed through the spreading of malignant cells from other parts of the body to the brain. One of the most frequent primary tumors is Glioma [2]. It affects not only the glial cells of the brain, but it invades also the surrounding tissues. The high-grade glioma (HGG) or glioblastoma (GBM) is the most common and aggressive type with a median survival rate of one to two years [3]. Another slower-growing low-grade glioma (LGG) such as astrocytoma, has slightly longer survival time. Treatment methods such as radiotherapy and chemotherapy may be used to destroy the tumor cells that cannot be physically resected or to slow their growth.

Therefore, neurosurgery still presents the initial and, in some cases, the only therapy for many brain tumors [4]. However, modern surgical treatment of brain tumors faces the most challenging practice conditions because of the nature and structure of the brain. In addition, distinguishing tumor tissue from normal brain parenchyma is difficult for neurosurgeons based on visual inspection alone [5]. Magnetic resonance imaging (MRI) is widely used as a common choice for diagnosing and evaluating the intraoperative treatment response of brain tumors [6]. Furthermore, MRI provides detailed images of the brain tumor cellularity, vascularity, and blood-brain barrier using different produced multimodal protocols such as T1-weighted, T2-weighted, and T2-FLAIR images. These various images provide information to neurosurgeons and can be valuable in diagnostics. However, interpreting these data during neurosurgery is a very challenging task and an appropriate visualization of lesion structure apart from healthy brain tissues is crucial [7].

---

* Corresponding author. Tel.: +49 (0) 7121 271 4007; Fax: +49 (0) 7121 271 4032.
*E-mail address:* Ramy.Zeineldin@Reutlingen-University.DE

Manual segmentation of the brain tumor is a vital procedure and needs a group of clinical experts to accurately define the location and the type of the tumor. Moreover, the process of lesion localization is very labor-based and highly dependent on the physicians' experience, skills, and their slice-by-slice decisions. Alternatively, automated computer-based segmentation methods present a good solution to save the surgeon's time and to provide reliable and accurate results, while reducing the exerted efforts of experienced physicians to accomplish the procedures of diagnosis or evaluation for every single patient [8]. Formerly, numerous machine learning algorithms were developed for the segmentation of normal and abnormal brain tissues using MRI images [9]. However, choosing features that enable this operation to be fully automated is very challenging and requires a combination of computer engineering and medical expertise. Therefore, classical approaches depend heavily on the applied application and do not generalize well. Nevertheless, developing fully automated brain tumor methods is still challenging task, because malignant areas varied in terms of shape, size, and localization, and they can only be defined through the intensity changes relative to surrounding healthy cells.

Recently, deep learning becomes an attractive field of machine learning that outperforms traditional computer vision algorithms in a wide range of applications such as object detection [10], semantic segmentation [11] as well as other applications such as navigation guidance [12]. Convolutional Neural Networks (CNNs) have proved during the ImageNet Large-Scale Visual Recognition Challenge (ILSVRC) [13] their ability to accurately detect and localize different types of objects. In 2012, an advanced pre-trained CNN model called AlexNet [14] showed the best results in the image classification challenge. Then, other CNN models have dominated the ILSVRC competition; namely Visual Geometry Group Network (VGGNet) [15], Residual Neural Network (ResNet) [16], Dense Convolutional Network (DenseNet) [17], Xception [18], MobileNet [19], NASNet [20], and MobileNetV2 [21]. Moreover, CNN methods have been applied to perform MRI tumor segmentation [22,23].

Semantic segmentation is currently one of the most important tasks in the field of computer vision towards complete scene understanding. Early approaches of applying semantic segmentation in the medical field use patch-wise image classification [24]. However, it suffers from two main problems: First, the training patches are much larger than the training samples, which require a higher number of computation cycles resulting in a large running time consumption. Second, the segmentation accuracy depends heavily on the appropriate size of patches. Accordingly, new network architecture was introduced, refer to Fig. 1, which is able to solve these problems by using two main paths: a contracting path (or encoder) and an expansive path (or decoder) [25]. The encoder is typically a CNN consisting of consecutive two 3x3 convolutional layers, each followed by a rectified linear unit (ReLU) and 2x2 spatial max pooling. Contrarywise, the decoder aims at upsampling the resultant feature map using deconvolution layers followed by 2x2 up-convolution, a concatenation layer with the corresponding down-sampled layer from the encoder, two 3x3 convolutions, and a ReLU. Finally, the upsampled features are then directed to a 1x1 convolution layer to output the final segmentation map. Remarkably, the networks are able to achieve precise segmentation results using only few training images with the help of data augmentation [26]. Furthermore, the tiling strategy allows the model to employ high-resolution images in the training stage with low GPU memory requirements.

This study aims at developing a new fully automated MRI brain tumor segmentation based on modified U-Net models, including the following contributions:
- Presenting the developed modular design of DeepSeg to include new segmentation architectures for the FLAIR modal brain MRI.

- Proposing a generic modular architecture of the brain tumor segmentation with two elements: feature extraction and image expanding paths, in order to support applying different deep neural network models successfully.
- A detailed ablation study of the state-of-the-art deep learning models highlighting the computational performance during training and prediction processes.
- Validating the proof of concept to apply various deep learning models for assisting the clinical procedures of the brain tumor surgery using FLAIR modality.

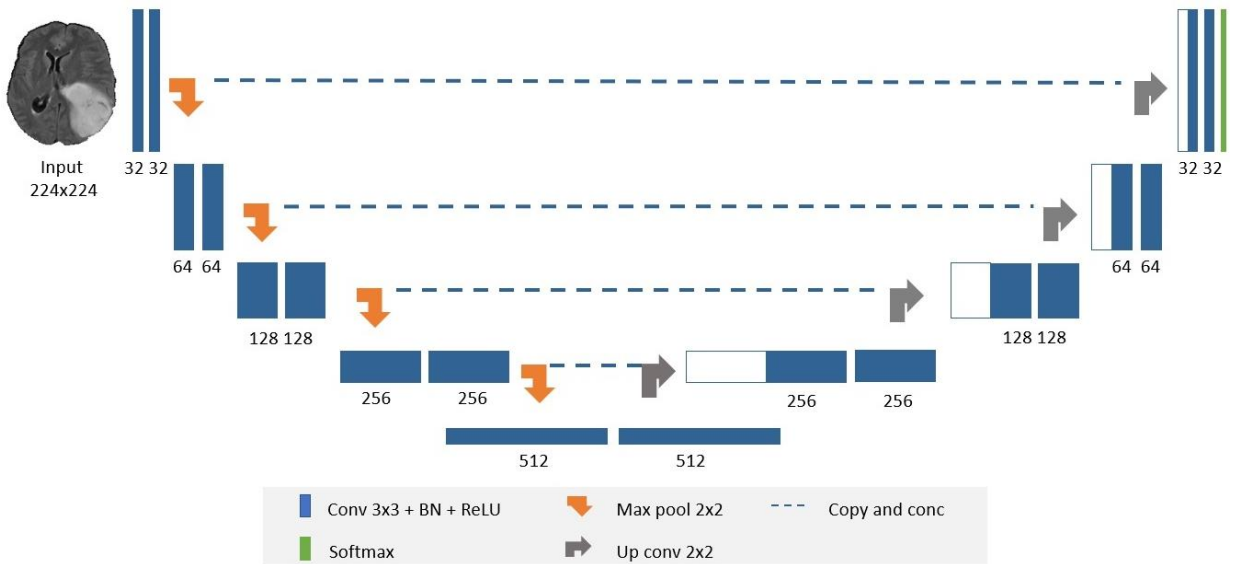

**Fig. 1** Modified U-Net network consists of convolutional blocks (*blue boxes*), maximum pooling (*orange arrows*), upsampling (*grey arrows*), and softmax output (*green block*).

## Methods

DeepSeg is a generic decoupled framework for automatic tumor segmentation, as shown in Fig. 2. Thanks to the basic U-Net structure [27], it consists of two main parts: a feature extractor (or encoder) part and an image upscaling (or decoder) part. This universal design has two main advantages: First, it allows the extensibility of the system, i.e. different encoders and decoders can be added easily. Moreover, a discriminative comparison between the various proposed models can be done straightfowardly. In the following, the proposed architecture is described in detail.

### Feature extractor

The modified U-Net encoder has been implemented by using advances in CNNs including dropout and batch normalization (BN) [28,29]. In addition, state-of-the-art deep neural network architectures are integrated into our benchmarking system to extract the feature map. These models are utilized to achieve better performance than the obtained results in the ILSVRC competition until now [13]. Apparently, every proposed model has its own set of parameters and computational resources requirements, as described below.

**VGGNet** [15] is proposed by the Visual Geometry Group from the University of Oxford and is the winner of the ILSVRC 2014 in the localization task. It is chosen to be the baseline model because of its simplicity, consisting only of small 3x3 convolutional layers and max-pooling layers for the downsampling process followed by two fully connected layers for feature extraction.

In fact, increasing the neural network layer would increase the accuracy of the training phase, however, there is a significant problem with this approach; for example, vanishing gradients [30] cause the neural network accuracy to saturate and then degrade rapidly. In ILSVRC 2015, a novel micro-architecture called **ResNet** [16] was introduced to solve this exploding behavior. The ResNet consists of residual blocks as shown in Fig. 3. (a), and each block consists of the original two convolutional layers in addition to a shortcut connection from the input of the first layer to the output of the second layer. By employing skip connections to the deep neural network, neither more additional parameters nor computational complexity are added to the network. Owing to this approach, they are able to train up to 152-layer deep neural network while maintaining a lower complexity than the above VGG models.

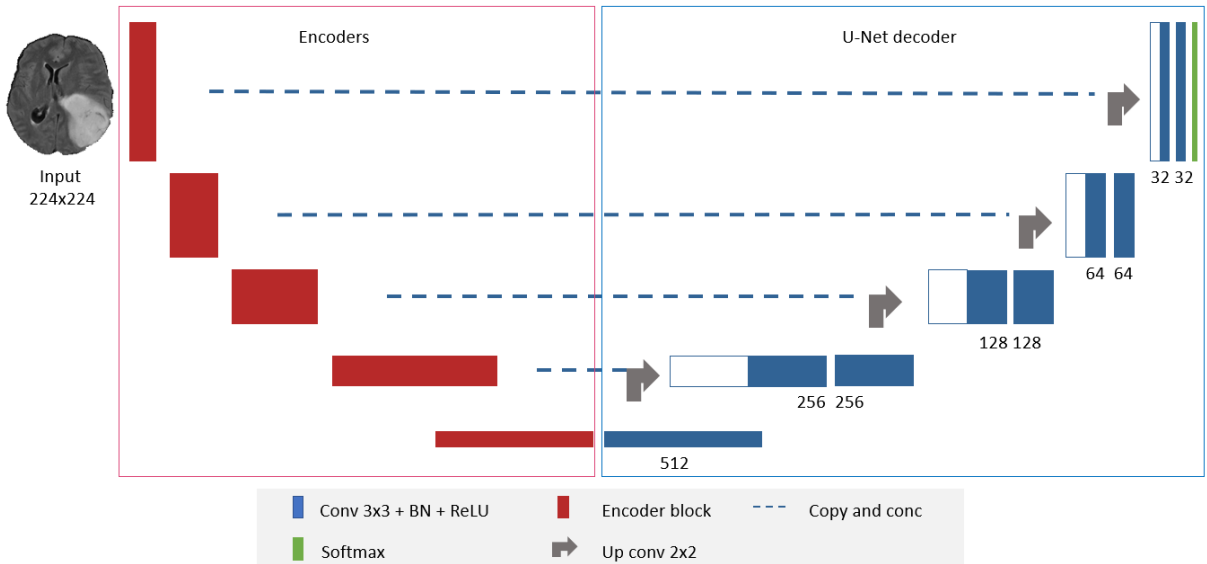

**Fig. 2** DeepSeg architecture for using different feature extractor models of MRI brain tumors.

**DenseNet** [17] uses the feature map of the preceding layers as inputs into all the following layers, as depicted in Fig. 3. (b). This type of deep neural network model has L(L+1)/2 connections for a CNN with L layers, whereas traditional networks would have only L connections. Remarkably, they are able to achieve additional improvements such as a smaller number of parameters besides the ability to scale the network to hundreds of layers.

**Xception** presents an extreme version of the Inception network [18]. The Inception model aimed at improving the utilization of the computing resources within the neural network through special modules. Each inception module is a multi-level feature extractor by stacking 1x1 and 3x3 convolutions beside each other in the same layer rather than using only one convolutional layer. The Xception, as shown in Fig. 3. (c), achieved a slightly better result than Inception models on ImageNet, however, it showed superior improvement when the used dataset becomes larger.

Google presented **MobileNet** [19] in 2017 as an efficient light-weight network for mobile application, as presented in Fig. 3. (d). Additionally, the BN is applied after each convolution followed by a ReLU activation. Then **MobileNetV2** [21] is introduced, which enhanced the state of the art performance of mobile models based on inverted residual blocks as shown in Fig. 3. (e). These bottleneck blocks are similar to the residual block of ResNet where the input of the block is added to the output of the narrow layers. ReLU6 is also utilized, because of its robustness in low-precision computation, to remove the non-linearities in the bottleneck layers. Although MobileNetV2 shows a similar performance to the previous MobileNet, it uses only 2.5 times fewer operations than the first version.

Google Brain introduced **NASNet** [20] to obtain state-of-the-art segmentation results with relatively smaller model size. The basic architecture of NASNet is made up of two main repeated blocks namely Normal Cell and Reduction Cell. The first type is consisting of convolutional layers with output features of the same dimensions, and the height and width of the other type's output are reduced by a stride of 2. ScheduledDropPath is also presented to make the model generalize well, where each path in the cell can be dropped with an increased probability over the training sequence.

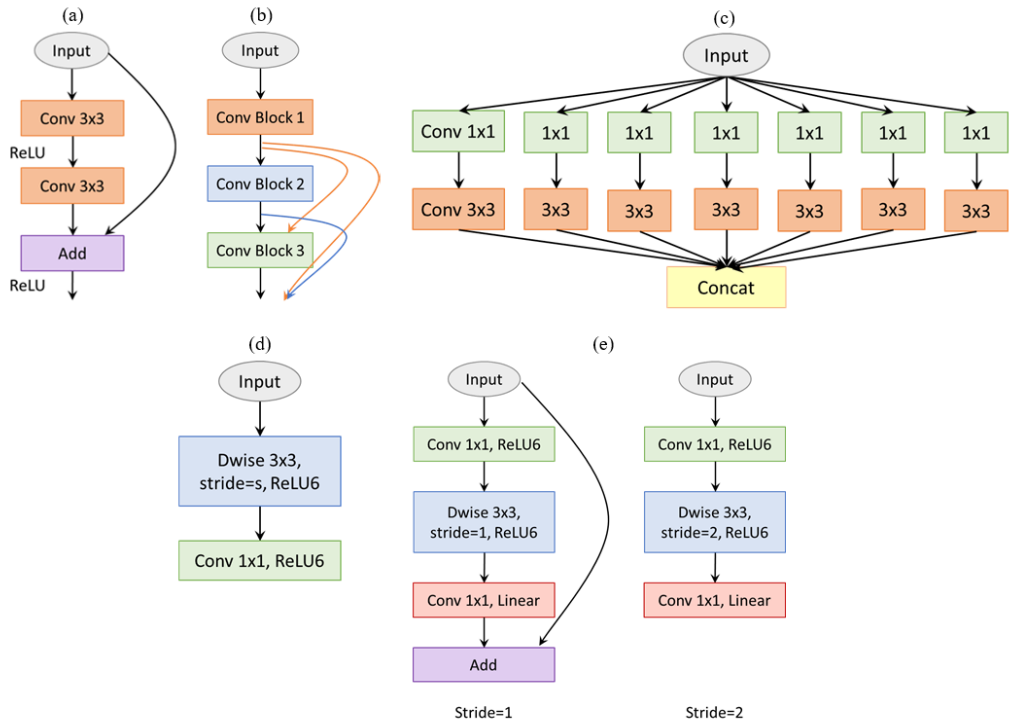

**Fig. 3** Comparison of the basic blocks for different feature extractors. (a) the residual block of ResNet; (b) a 3-layer dense block; (c) an extreme module of Inception (Xception module); (d) a depth-wise based module of MobileNet; (e) MobileNetV2 blocks with two stride values.

**Image upscaling**

In semantic segmentation, it is very crucial to use both semantic and spatial information so that the neural network can perform well. Hence, the decoder should recover the missing spatial information to get the full resolution segmented map from the consequential encoded features. By skip connections (Fig. 1), U-Net can obtain the semantic feature map from the bottleneck and recombine it with higher resolution outputs from the encoder respectively.

Unlike standard U-Net decoder, some modifications were incorporated for further exceptional segmentation results. Firstly, a BN layer is applied between each convolution and ReLU to make each layer learn independently from other layers and thus contribute to faster learning. Additionally, a smaller filter size of 32 as the base filter is selected and doubled at the following layers, in order to apply the full size as input rather than using patches or small regions of the input. Finally, the output of the network is passed into a softmax output layer which converts the output logics into a list of probability distributions.

**Data**

This study was performed using the FLAIR MRI data from the BraTS 2019 challenge [31]. Although T1KM is the standard imaging for Glioma, FLAIR is becoming increasingly relevant in the case of malignant tumors, since there is a trend to also resect the FLAIR positive areas [32]. Moreover, the advantages of FLAIR images in the brain surgery of low-grade gliomas (LGG) have been investigated by our clinical partners in [6].

BraTS dataset contains multi-institutional pre-operative MRI of 336 heterogeneous (in shape, appearance, size, and texture) Glioma patients (259 HGG and 76 LGG). Each patient has four multimodal scans: native T1-weighted, post-contrasted T1-weighted, T2-weighted, and T2-FLAIR. MRI data were acquired with various clinical protocols and different scanners from 19 institutions. The manual segmentation of the data was done by experienced neuro-radiologists, from 1 to 4, following the same annotation procedure. After that, the MRI scans are resampled and interpolated to the same resolution 1 mm$^3$. Fig. 4 displays the provided segmented labels: the necrotic and non-enhancing tumor core (label 1), the peritumoral edema (label 2), and enhancing tumor (label 4).

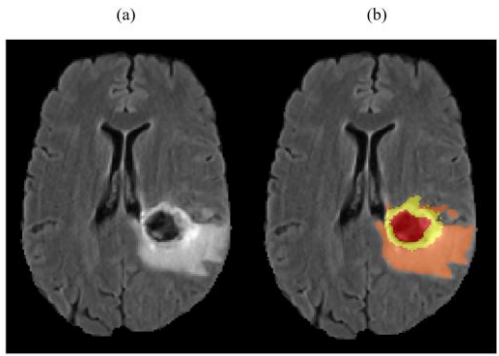

**Fig. 4** (a) T2-FLAIR image; (b) brain tumor structures including non-enhancing core and necrotic (*red*); solid core (*yellow*) and edema (*orange*).

Since MRI scans come from different sources, protocols, and institutions, the training images may suffer from bias field noise, which can be defined as undesirable artifacts that arise during the process of image acquisition. To eliminate these effects, the improved N3 bias correction tool [33] is used for performing image-wise normalization and bias correction. Then, a data normalization for each slice of FLAIR MRI scans is applied by subtracting the mean of each slice and dividing by its standard deviation.

**Table 1** List of the applied data augmentation methods.

| Methods | Parameters |
|---|---|
| Flip horizontally | 20% of all images |
| Flip vertically | 20% of all images |
| Scale | ±20% on both horizontal and vertical direction |
| Translation | ±20% on both horizontal and vertical direction |
| Rotation | ±25° |
| Shear | ±8° |
| Elastic transformation | α=720, σ=24 |

Providing we are training large neural networks using limited training data, some precautions should be taken to prevent the problem of overfitting. One of them is data augmentation, which is the process of creating new artificial training data from the original one in order to improve the model performance by making the model generalize well to the new testing data. In this study, a set of simple on-the-fly data augmentation methods is applied (as listed in Table 1) by horizontal and vertical flipping, rotation, scaling, shearing, and shift. Unfortunately, these simple methods are not enough to get sufficient immune training data, therefore more complex methods are also introduced such as elastic distortion corresponding to uncontrolled noise of MRI sensors, where σ is the elasticity coefficient and α is the multiplying factor of the displacement fields which controls the intensity of deformation. Figure 5 shows some examples of the applied augmentation techniques.

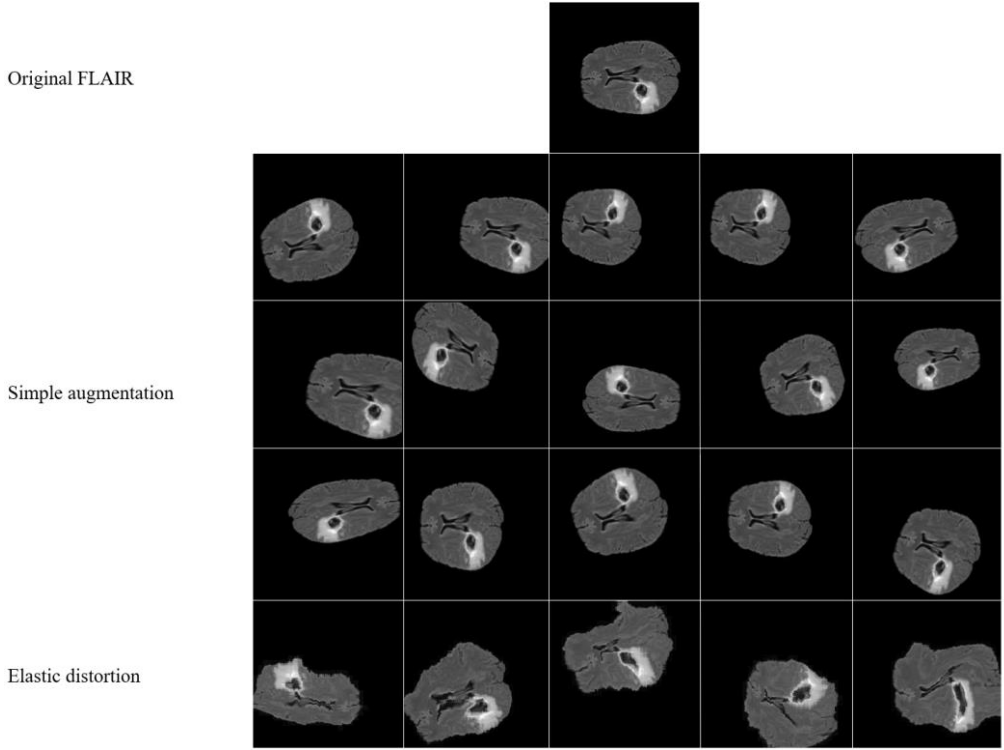

Original FLAIR

Simple augmentation

Elastic distortion

**Fig. 5** Random augmented image transformation. The first row shows the original image. Next three rows present horizontal and vertical flipping, scaling, translation, rotation, and shearing methods. The elastic transformation is presented in the last row.

## Experiments

### Experimental setup

The proposed methods of this study were run on AMD Ryzen 2920X (32M Cache, 3.50 GHz) CPU with a single 11GB NVIDIA RTX 2080Ti GPU. Proposed models are implemented in Python using Keras library and Tensor Flow backend. Experiments are done using FLAIR MRI sequences with a resolution of 224x224 in order to use all the proposed feature extractor networks. All networks are trained for 35 epochs and a batch size of 16. During the training process, spatial dropout with a rate of 0.5 was used after the feature extractor path. This is a simple type of regularization to ensure that the neural networks generalize well without overfitting the training dataset. Adam optimizer [34] has been applied with a learning rate of 0.00001. Nevertheless, the BraTS dataset suffers from a data imbalance problem where the tumor pixels are less than 2% and the healthy pixels are mostly 98% of the whole dataset. To solve this problem in two steps, firstly, the models were trained on the brain sequences and ignored the empty slices, secondly, a weighted cross-entropy loss for each class was used to pay more attention to the malignant labels than the background as defined by

$$L = -\sum_{n=1}^{N} y_c \, log(p_c) * w \tag{1}$$

where $N$ is the number of classes including the background and the tumor cells in this study, $y_c$ represents the true labels for the $n^{th}$ class, $p_c$ is the predicted softmax output for those true labels, and $w$ is the proposed weight map of (0.05, 0.95) to focus on the tumor pixels rather than the background. For the evaluation of our segmentation results, four metrics namely; dice similarity coefficient (DSC), sensitivity, specificity and the Hausdorff distance (HD) are computed. DSC score calculates the overlap of the segmented region and the ground truth y and is applied to the network softmax predictions p as follows:

$$DSC = \frac{2 * \sum yp + \varepsilon}{\sum y + \sum p + \varepsilon} \qquad (2)$$

Note that $\varepsilon$ is the smooth parameter to make the dice function differentiable. This dice overlap can take values from 0 (represents lower overlap) to 1 (indicates a full overlap). Specificity and sensitivity are given by:

$$Specificity = \frac{TN}{TN + FP} \qquad (3)$$

$$Sensitivity = \frac{TP}{TP + FN} \qquad (4)$$

where true positives (TP) and false positives (FP) refer to the number of retrieved points that are correct/incorrect, and similarly for true and false negatives, TN and FN, respectively.

Dice, Sensitivity, and Sensitivity metrics are measures of pixel-based overlap between the ground truth and the predicted segmentations. In addition, the HD gives the largest distance of the segmentation set to the nearest point in the truth set, as defined by

$$HD(S, T) = max\{h(P, T), \ h(T, P)\} \qquad (5)$$

with

$$h(S, T) = max_{s \in S}\{min_{t \in T}\{d(s, t)\}\} \qquad (6)$$

where the shortest Euclidian distance $d(s, t)$ is calculated for every point $s$ of the segmentation set $S$, with respect to the ground truth point $t$ in the image.

**Ablation Study**

Thanks to the DeepSeg framework, several methods were analyzed and compared simultaneously. Table 2 illustrates different characteristics of these automated methods with the corresponding computational times. Training and prediction times present the average estimated time of applying each algorithm about 35 times during the training and validation respectively. These tests showed that MobileNet encoder requires smallest resources with only 22 MB memory and roughly 5.6 thousand parameters. It is worth mentioning that MobileNet and MobileNetV2 are mainly developed for mobile and embedded applications where the hardware resources are limited. Likewise, U-Net, modified U-Net, and NASNet consumes a small amount of memory of 30 MB, 30 MB and 37 MB respectively. Obviously, there is a proportional relationship between the number of parameters and the demanded memory. In contrary, ResNet model consumes the largest amount of memory of 118 MB, which is not considered a problem since modern GPUs possess a memory of several Gigabytes. Other models such as DenseNet, VGGNet, and XCeption are located in the middle level of memory consumption of 51 MB, 71 MB, and 103 MB, respectively.

Moreover, the number of layers has a significant influence on both the training and prediction time. For instance, the training time of one epoch using U-Net with the smallest number of layers (39 layers), is 381 seconds and the prediction time is just 1.1 seconds. But the NASNet model with 818 layers requires 684 seconds for one epoch to train and the prediction of one patient took 4.4 seconds. Nevertheless, this is not the general rule since modified U-Net, MobileNet, and MobileNetV2 share the second place with a training time of 385 seconds even though they have various numbers of layers of 74, 129, and 202, respectively. The main reason is the internal building architecture of MobileNet variants which is developed for smartphone devices.

**Table 2** A comparative performance of the employed models. Average computational times for each encoder of 35 results during training and validation phases.

| Encoder | Size (MB) | Training Time (sec) | Prediction Time (sec) | Parameters | Layers |
|---|---|---|---|---|---|
| U-NET | 30 | **381** | **1.1** | 7760642 | **39** |
| Modified U-NET | 30 | 385 | 1.3 | 7763050 | 74 |
| VGGNet | 71 | 540 | 1.6 | 18540938 | 56 |
| ResNet | 118 | 446 | 2.3 | 30546458 | 223 |
| DenseNet | 51 | 482 | 3.2 | 12947674 | 474 |
| XCeption | 103 | 580 | 1.9 | 26769602 | 184 |
| MobileNet | 30 | 385 | 1.5 | 7590746 | 129 |
| NASNet | 37 | 684 | 4.4 | 8652846 | 818 |
| MobileNetV2 | **22** | 386 | 1.8 | **5591770** | 202 |

## Segmentation results

The DeepSeg framework consists of several automated feature extractors in addition to an image expanding path. The corresponding evaluation results have been obtained by running two-fold cross-validation on the 336 training cases of the BraTS 2019 dataset divided as follows: 270 cases for training and 66 for validation. Table 3 summarizes the comparison and the overall measurement results of all tested methods.

**Table 3** Mean DSC, Sensitivity, Specificity and Hausdorff distance scores of testing different encoders on BraTS 2019 training data.

| Encoder | DSC | Sensitivity | Specificity | HD |
|---|---|---|---|---|
| U-NET | 0.809 | 0.799 | **0.998** | 12.926 |
| Modified U-NET | 0.814 | 0.783 | **0.999** | 13.341 |
| VGGNet | **0.837** | 0.819 | **0.998** | 12.633 |
| ResNet | 0.811 | 0.789 | **0.998** | 13.652 |
| DenseNet | **0.839** | 0.827 | **0.998** | 13.156 |
| XCeption | **0.839** | **0.856** | **0.998** | 11.337 |
| MobileNet | **0.835** | 0.843 | **0.998** | **10.924** |
| NASNet | 0.834 | 0.826 | **0.998** | 12.608 |
| MobileNetV2 | 0.827 | 0.822 | **0.998** | 12.029 |

The proposed deep learning architectures were able to accurately detect tumor regions in the validation set with mean DSC scores ranging from 0.809 to 0.839, while the mean dice score of the expert's annotation for the whole tumor core is about 0.85 as reported in [35]. Although statistical analysis of results is relatively close or identical (like Specificity), these results give an important indication that fully automated deep learning models maybe utilized in the task of brain tumor segmentation. As illustrated in Table 3, The DenseNet, Xception, VGGNet, and MobileNet encoders achieved the best DSC scores of 0.839, 0.839, 0.837, and 0.835, respectively. Although the Xception encoder showed the best value for the sensitivity of 0.856 with approximately 7% better than the original U-Net model, it achieved the same value of the specificity. This result confirms that point-based approaches are not enough for evaluating brain tumor segmentation method. Therefore, the HD measurements were applied to verify both the best accuracy and performance among all tested deep encoders. The MobileNet showed the shortest HD value of 10.924.

Figures 6 and 7 show segmentation results for the proposed architectures generated from the validation set (67 cases). In both figures, the first row indicates the FLAIR images in gray color mode and the manual ground truth segmentations are shown in the second row. In the following rows, segmentation results of different automated methods are presented. It can be observed that segmented tumor boundaries (indicated in red) from proposed encoders are very similar to the manual annotation even when the lesion region is heterogeneous in shape, volume, and texture. For instance, a small-sized tumor in case TCIA12_470_1 was accurately segmented by proposed methods, however, when the heterogeneity of malignant cells increases, the performance varied remarkably. This is clear in TCIA10_103_1 case since some encoders such as U-Net, VGGNet, MobileNet tends to over-segment the tumor area, while modified U-Net, ResNet, Xception, NASNet, and MobileNetV2 tend to under-segment. This result showed superior accuracy of the Xception, DenseNet encoders compared to other tested architectures for the most difficult tumor segmentation case, e.g. the case of TCIA10_351_1. Although the DenseNet encoder provided a lower score of tumor segmentation result in the case of TCIA10_490_1, it is valid and clinically accepted. However, other encoders such as U-Net and NASNet are failed to give accepted segmentation results.

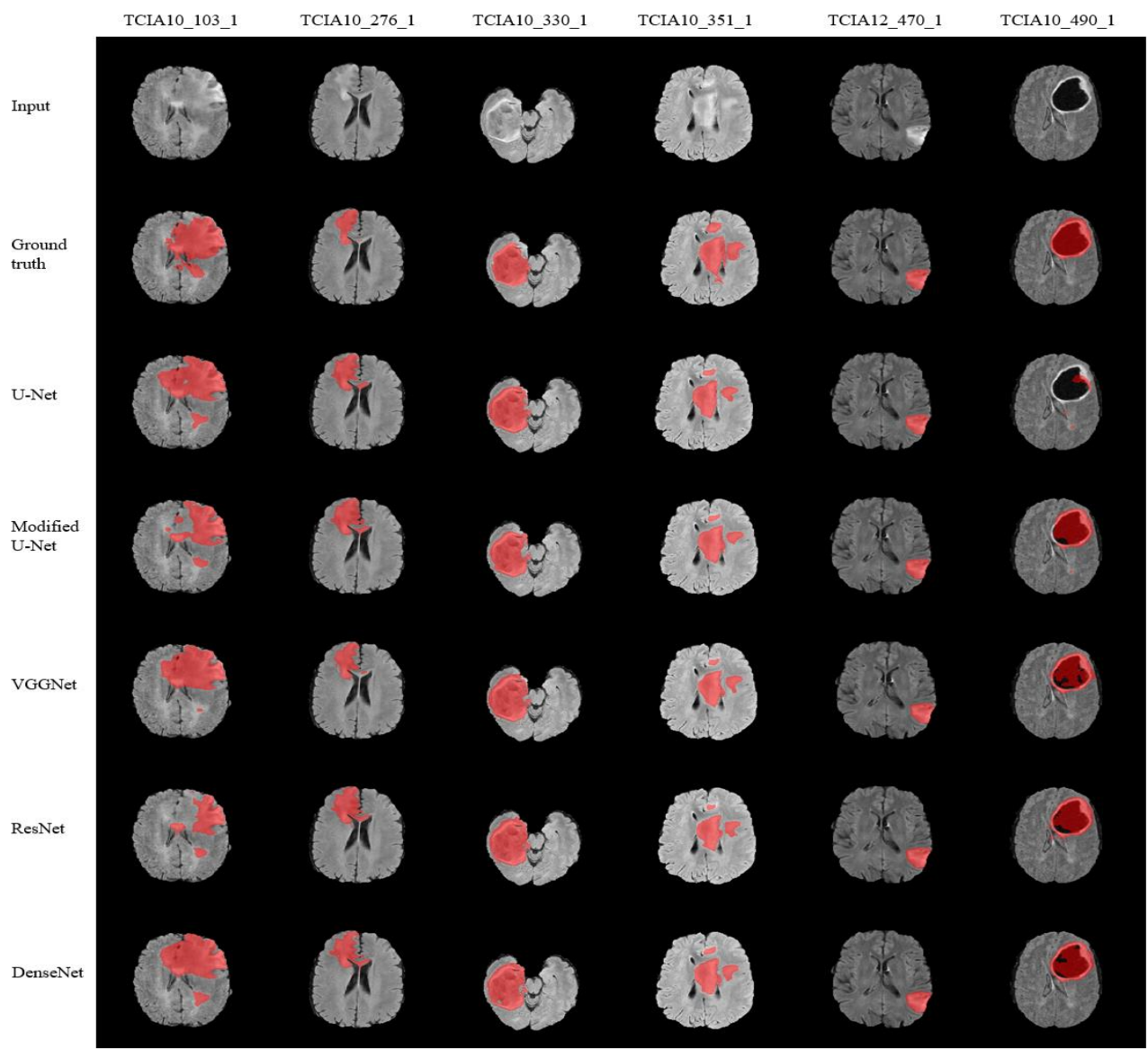

**Fig. 6** Brain tumor segmentation results. T2-FLAIR, ground truth and output of Original U-Net, Modified U-Net, VGGNet, ResNet, and DenseNet. Tumor regions are indicated in red.

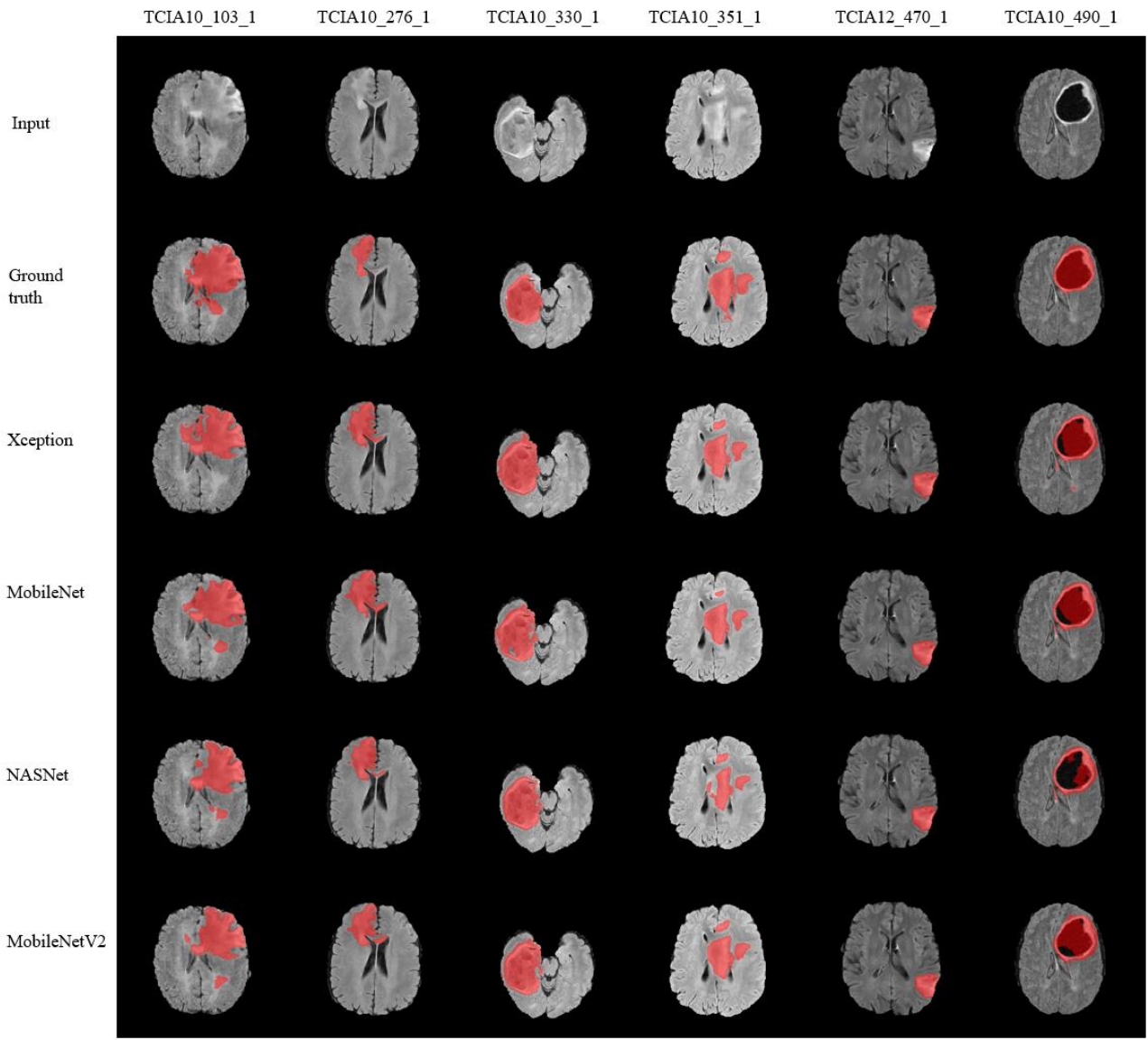

**Fig. 7** Brain tumor segmentation results. T2-FLAIR, ground truth and output of Xception, MobileNet, NASNet, and MobileNetV2.

## Evaluation

For consistency with other publications, the proposed architectures have been also tested on the validation datasets of BraTS 2019 (125 cases). Table 4 presents the compared scores of mean dice similarity coefficient, sensitivity, specificity, and HD, similar to the online evaluation platform (https://ipp.cbica.upenn.edu/). These results showed that the proposed models are robust and able to deal with MRI segmentation task. In Table 4, the DenseNet architecture outperformed other models with respect to the DSC (0.841) as well as in the training set, however, it ranked second with a HD (10.595) which is clinically accepted. The dice metrics and HD are the most important measurements when evaluating and comparing among deep learning models, because they show the percentage of the overlapping between ground truth segmentation and predictions. In contrast, the lack of false positives indicated high values of both specificity and sensitivity, which may not precisely reflect the actual performance.

In Table 4, the top-score team "Questionmarks" wins currently the first place of segmentation task rankings from the BraTS 2019 challenge. The performance of our proposed methods does not exceed this score. However, our segmentation results are still clinically accepted due to the following reasons: First, the DeepSeg methods are trained using only FLAIR MRI data dissimilar to participating teams in BraTS 2019, because multi-MRI modalities are not always applicable and sometimes would be unfeasible in clinical experiments. Finally, the online evaluation system presents unranked leaderboard and the calculated score is an average of all the submissions made by the team.

**Table 4** Mean DSC, Sensitivity and Specificity scores of applied models on BraTS 2019 validation data.

| Encoder | DSC | Sensitivity | Specificity | HD |
|---|---|---|---|---|
| U-NET | 0.813 | 0.841 | 0.987 | 19.747 |
| Modified U-NET | 0.820 | 0.853 | 0.987 | 12.014 |
| VGGNet | 0.829 | 0.837 | 0.990 | 9.756 |
| ResNet | 0.823 | 0.832 | 0.990 | 10.005 |
| DenseNet | 0.841 | 0.860 | 0.989 | 10.595 |
| XCeption | 0.834 | 0.865 | 0.988 | 12.571 |
| MobileNet | 0.830 | 0.855 | 0.989 | 11.696 |
| NASNet | 0.830 | 0.861 | 0.988 | 11.673 |
| MobileNetV2 | 0.822 | 0.854 | 0.988 | 13.894 |
| Questionmarks | 0.909 | 0.924 | 0.994 | NA |

## Conclusions

This study demonstrated the feasibility of employing deep learning approaches for assisting the procedures of brain surgery. The DeepSeg framework is developed successfully for fully automated segmenting brain tumors in MR FLAIR images, based on different architectures of deep CNN models. Moreover, the findings of this comparative study have been validated using the BraTS online evaluation platform, as illustrated in Table 4.

Currently, we are working on extending the validation of our DeepSeg framework by adding more image datasets from other different MRI modalities such as T1- and T2-weighted to verify its potential impact on the planning procedures of the brain tumor surgery, with our clinical partners at the Department of Neurosurgery, University of Ulm. Furthermore, processing 3-D convolutions with like atrous spatial pyramid pooling (ASPP) [36] over all slices will advance the DeepSeg framework to cover the clinical requirements for accurate segmentation of brain tumors during MRI-guided interventions.

**Acknowledgement** The corresponding author is funded by the German Academic Exchange Service (DAAD) under scholarship No. 91705803.

**Compliance with Ethical Standards**

**Conflict of Interest** The authors have no conflict of interest to disclose.

**Ethical Approval** This article does not contain any studies with human participants or animals performed by any of the authors.

**Informed Consent** This article does not contain patient data.